\documentclass[prb, preprintnumbers, amsmath, amssymb, superscriptaddress, twocolumn]{revtex4}
\usepackage{graphicx, color, dcolumn, bm, natbib, amssymb, amsmath}
\usepackage[final=true]{hyperref}

\begin{document}

\title{Magnetodipolar interaction and quasiparticles delocalization in disordered quantum magnets}

\author{Oleg I. Utesov}
\affiliation{Saint Petersburg State University, Ulyanovskaya 1, Saint-Petersburg, 198504, Russia}
\affiliation{Petersburg Nuclear Physics Institute, Gatchina, St-Petersburg, 188300, Russia}
\affiliation{St. Petersburg School of Physics, Mathematics, and Computer Science, HSE University, 190008 St. Petersburg, Russia}

\newcommand{\m}{\mathbf}
\newcommand{\h}{\hat}
\newcommand{\ve}{\varepsilon}
\renewcommand{\c}{\cdot}
\renewcommand{\o}{\omega}
\renewcommand{\t}{\theta}
\newcommand{\be}{\begin{eqnarray}}
\newcommand{\ee}{\end{eqnarray}}
\newcommand{\nn}{\nonumber}

\begin{abstract}

It is well-known that disordered quantum magnets usually host localized elementary excitations in gapped phases. Here we show that long-range magnetodipolar interaction can lead to quasiparticles delocalization in the three-dimensional case, similarly to the phenomenon previously described for localized vibrational modes in disordered solids. Employing renormalization group-like ideas, the average eigenmode occupation number is found to scale with the system size, which also leads to sub-diffusive dynamical properties of the system.

\end{abstract}

\maketitle

\section{Introduction}

Quantum magnets establish themselves as a fertile playground for many concepts of condensed matter physics, for instance, quantum phase transitions~\cite{sachdev2011}. Both ordered~\cite{giamarchi2008bose} and disordered~\cite{zheludev2013dirty} cases can be studied using magnetic insulators such as spin-dimer systems, where in the absence of external field pairs of spins form singlets, and the triplet excitations (triplons) are well separated from the ground state by the gap~\cite{sachdev1990}. In the disorder-less case corresponding phase transitions between gapped and ordered phases can be usually described as Bose-Einstein condensation of elementary excitations~\cite{batyev1984,batyev1985,tlcucl3,paduan2004,zapf2005bose,yin2008,kofu2009,utesov2014jmmm}. Importantly, when quenched disorder is introduced (by, e.g., chemical substitution of non-magnetic ions, which change magnetic interaction parameters locally~\cite{manaka2001,oosawa2002,hong2010,yu2012bose,huvonen2012,povarov2015dynamics,gazizulina2017}), intermediate Bose glass phase should emerge between gapped and ordered phases~\cite{fisher1989,theorem}. It was indeed observed experimentally in bromine-doped IPA-CuCl$_3$~\cite{hong2010} and in bromine-doped dichloro-tetrakis-thiourea-nickel (DTN) in Ref.~\cite{yu2012bose}.

For physics of disordered bosonic systems, as well as for strongly disordered metals, the concept of quasiparticles localization~\cite{Anderson1958} is important. The transition from gapped phase to Bose glass was shown to be of the Griffiths type~\cite{Gurarie2009,Griffiths1969}, which rely on low-energy elementary excitations localization. In quantum magnets, corresponding phenomenon was addressed theoretically in Refs.~\cite{utesov2014,utesov2021SCTMA}. However, there are several mechanisms that can lead to quasiparticles delocalization and further complicate discussed above behavior.  The first mechanism is due to quasiparticles interaction, which can be called many-body delocalization (see Ref.~\cite{gorniy2017} and references therein). The second one relies on usually weak long-range interactions, which can provide effective hopping between localized states, see Refs.~\cite{levitov1989,levitov1990}. In these papers, it was shown that localized phonons in disordered media (vibrons) are delocalized due to long-range electric dipole interaction, no matter how weak it is.

Here we report a similar delocalization effect in disordered three-dimensional quantum magnets, which arises due to mangetodipolar interaction. Localized states can be represented as magnetic dipoles on the top of a non-magnetic background and dipolar forces allow for excitation to hop between the states in resonance, which number diverges in the thermodynamic limit. We derive the main properties of the delocalized states' critical wave functions and system dynamics using simple scaling arguments.

\section{Theory}

\subsection{Tunneling Hamiltonian}

Theoretical consideration below is based on properties of a simple model of two coupled quantum states $| 1 \rangle, \, | 2 \rangle$ (or two coupled oscillators considered in Refs.~\cite{levitov1989,levitov1990} with slightly, but not essentially, different properties), which Hamiltonian in the matrix form reads
\be
  \mathcal{H} = \left(
                  \begin{array}{cc}
                    E_1 & V_{12} \\
                    V^*_{12} & E_2 \\
                  \end{array}
                \right).
\ee
The states hybridization is substantially dependent on the ``tunneling'' matrix element $ V = |V_{12}|$ to $\Delta E = |E_1-E_2|$ ratio. In the off-resonance case, $ V \ll \Delta E$, the wave-functions admixture is small, being of the order of $V/\Delta E$. In the opposite case of $\Delta E \lesssim V$  (resonance condition), the wave-functions are essentially hybridized and their characteristic oscillation time between two basis states is $\sim 1/V$.

So, we arrive at a conclusion, that when discussing delocalization caused by weak magnetodipolar interaction only close in energy states should be taken into account (it is also true for diffusion in Anderson model, see Ref.~\cite{Thouless1970}).

\subsection{Disordered magnetic systems}

As an example of systems for which our theory is applicable, we consider three-dimensional magnets with integer spin and large single-ion easy-plane anisotropy. The corresponding Hamiltonian reads
\be \label{ham1}
  \mathcal{H} = \sum_{i,j} J_{ij} \m{S}_i \cdot \m{S}_j + \sum_i \mathcal{D}_i (S^z_i)^2 - h \sum_i S^z_i.
\ee
Here $\mathcal{D}_i$ are anisotropy constants, which can vary from site to site (so-called ``diagonal disorder''), and $h = -g \mu_B H_z$ is the external magnetic field in energy units. In the disorder-less case ($\mathcal{D}_i \equiv \mathcal{D}$) the system is in gapped paramagnetic (PM) phase at $h<h_{C1}$ with all spins in state with $S^z=0$, and in gapped fully polarized (FP) phase at $h>h_{C2}$ with all  $S^z=1$, whereas at $h_{C1}<h<h_{C2}$ ordered canted antiferromagnetic phase appears\cite{kaganov1987}.

When the disorder of sufficient strength is introduced, localized states can emerge inside the elementary excitations gap of PM and FP phases (see Refs.~\cite{utesov2014, utesov2021SCTMA} for details). Below we assume that the localized states are distributed over some characteristic energy scale $\Delta \ve$.

Magnetodipolar interaction can be considered as a perturbation to the Hamiltonian~\eqref{ham1}, which reads
\begin{eqnarray}
 \label{hamD}
  \mathcal{H}_d &=& \frac12 \sum_{i,j} D^{\alpha \beta}_{ij} S^\alpha_i S^\beta_j,
\end{eqnarray}
where $\alpha, \beta$ denote Cartesian coordinates. The dipolar tensor is given by
\begin{equation}\label{dip1}
	 {\cal D}^{\alpha \beta}_{ij} = \omega_0 \frac{v_0}{4 \pi} \left( \frac{\delta_{\alpha\beta}}{R_{ij}^3} - \frac{3 R_{ij}^\alpha R_{ij}^\beta }{R_{ij}^5}\right),
\end{equation}
$v_0$ being the unit cell volume. The characteristic energy of the dipole interaction reads
\begin{equation}\label{dipen}
  \omega_0 = 4 \pi \frac{(g \mu_B)^2}{v_0}.
\end{equation}
which is usually of the order of $0.1 \div 0.5$~K. Despite this interaction being usually much weaker than the exchange coupling, it can be important because it is anisotropic and long-range. The latter property is crucial for the analysis below.

We proceed with the effect of dipolar interaction on the localized excitations taking S=1 for definiteness. In the PM phase we can use spin operators representation proposed in Ref.~\cite{sizanovBos}:
\be \label{spinrep1}
  S^+_i \approx \sqrt{2} (b^+_i + a_i), \, S^-_i \approx \sqrt{2} (b_i + a^+_i), \\
  S^z_i = b^+_i b_i - a^+_i a_i. \nn
\ee
Here two types of bosons are introduced (they represent $|\pm 1 \rangle$ states of each spin) and nonlinear terms in $S^\pm$ operators, which contribute only to quasiparticles interaction.

Despite localized states usually having some characteristic localization length, at small defects concentrations $n$ when considering long-range dipolar interaction we can neglect this length and assume that quasiparticles ``sit'' on the corresponding sites. Moreover, in the next section, it is shown that for the phenomenon of elementary excitations delocalization very long distances play the main role. Dipolar forces provide various contributions for excitations on sites $i$ and $j$, namely, $S^z_i S^z_j, S^\pm_i S^\pm_j, S^z_i S^\pm_j, S^\pm_i S^z_j$. Evidently, there are terms that allow excitation to hop between these sites, other terms are either of interaction-type or involving multi-particle states, which lie much higher in energy than the single-particle ones and can be safely neglected.

The same is true for localized states in the FP phase. Using standard Holstein-Primakoff spin operators representation~\cite{holstein}
\be \label{spinrep2}
  S^+_i \approx \sqrt{2S} a_i, \,  S^-_i \approx \sqrt{2S} a^+_i, \\
  S^z_i = S - a^+_i a_i, \nn
\ee
where spins are quantized with respect to magnetic field direction, we arrive at the same conclusion with the PM phase. Linear in bosonic operators terms lead to unimportant fine adjustment of local molecular fields, whereas quadratic terms provide hopping between impurity sites.

Finally, we note that for spin-dimer systems in the PM phase the ground state is the singlet one. So, it cannot be excited by $S^+_{i,1} + S^+_{i,2}$ (and similar) combinations of spin operators for a certain dimer, which should be put in dipolar forces Hamiltonian~\eqref{hamD}. To excite triplet state microscopic structure of dimer should be taken into account, however, corresponding terms will decrease with distance faster than $R^3_{ij}$ and do not lead to delocalization.

\begin{figure}
  \centering
  \includegraphics[width=8cm]{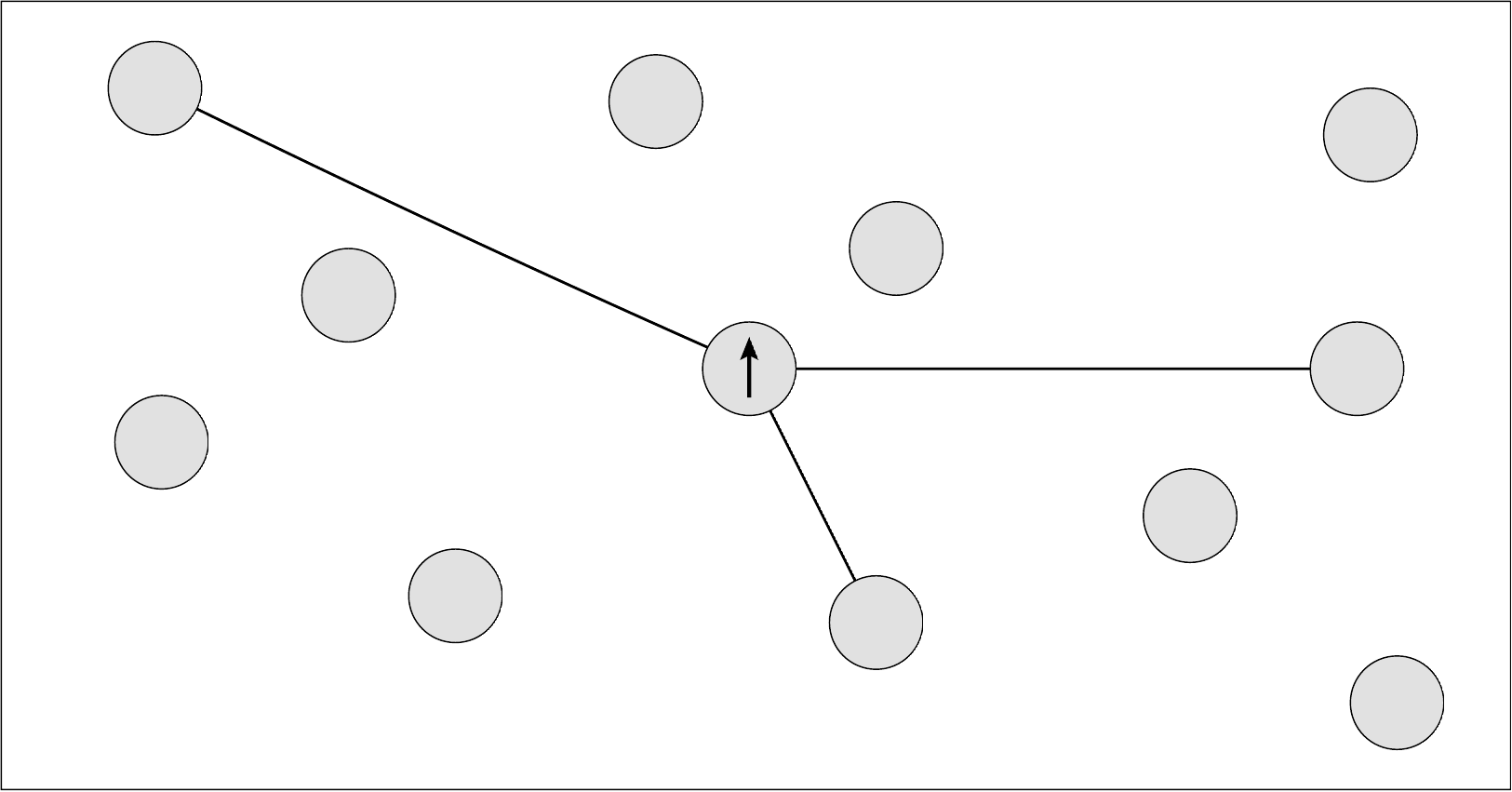}\\
  \caption{Illustration of the problem considered in this paper. Localized elementary excitation can hop between the bound on impurities states (which are shown using gray blobs) due to magnetodipolar interaction. Despite it being weak and resonances are rare, the long-range character of the dipolar forces leads to quasiparticles delocalization.  }\label{fig1}
\end{figure}

\subsection{Delocalization}

Now we consider a problem of a single particle wandering on a random network, where the sites with volume concentration $n$ are connected via dipolar-like interaction which for simplicity can be written as
\be \label{V1}
  V_{ij} = \frac{\lambda v_0}{4 \pi R^3_{ij}}.
\ee
Evidently, it provides the correct order of magnitude for hopping matrix elements if $\lambda \sim \omega_0$.

We can estimate the average number of resonances for one localized state as follows:
\be \label{nav}
  N = \int d^3 r \frac{n \lambda v_0}{4 \pi r^3 \Delta \ve} = \frac{n \lambda v_0}{\Delta \ve} \ln{\frac{R_0}{a_0}} \equiv u \ln{\frac{R_0}{a_0}}.
\ee
Here parameter $u= n \lambda v_0 /\Delta \ve \ll 1$, and the lower limit of integration is taken as a lattice parameter, so this equation is correct only with logarithmic accuracy. Moreover, in order to make the answer finite we manually introduce cut-off radius for dipolar forces $R_0$, which is infinite in reality (finite sample size $L$ can also serve as a cut-off parameter instead of $R_0$ in this formula). So, in the infinite sample (thermodynamic limit) the number of resonances diverges, which leads to delocalization of quasiparticles~\cite{Thouless1970}.

Next, we can take two localized states with \mbox{$ |E_i - E_j| = \Delta E$} which are in resonance. Then, the distance between them $R_{ij} \lesssim (\lambda v_0/\Delta E)^{1/3}$. If the third localized state is in resonance with both of them, it is easy to see that it cannot be further than $\sim R_{ij}$ from both of the sites $i$ and $j$, in other cases it can be only in resonance with one of these states (at large distances either $|E_i - E_k| \ll \Delta E$ or $|E_j - E_k| \ll \Delta E$). So, the probability to find the localized state that is in resonance with both states $|i\rangle$ and $|j\rangle$ can be estimated as
\be
  P_\triangle \sim n R^3_{ij} \frac{\Delta E}{\Delta \ve} \sim u \ll 1.
\ee
It means that the probability to find a triangle of bonds in resonance between three localized states is small. So, we make a conclusion, that the structure on which the quasiparticle has its random walk is a tree-like graph (without loops) where each vertex has in average $N$ edges (see Eq.~\eqref{nav}) at given $R_0$ or system size $L$ (for finite systems one should put $L$ instead of $R_0$ in Eq.~\eqref{nav}).

Based on the statements above, we can now analyze the behavior of the delocalized wave functions. To do it we use ideas of renormalization group theory: we assume that the number of sites participating in the wave function for the system size $L$ is $W(L)$ and see what happens when the size becomes $L + dL$. With logarithmic accuracy each of $W(L)$ vertexes acquire $u \ln{\left[(L+dL)/L\right]}$ new edges, so
\be
  W(L + dL) &=& W(L) +  W(L) u \ln{\frac{L+dL}{L}}    \nn \\
  \Longrightarrow \frac{d W}{W} &=& u \frac{d L}{L}. \label{rg1}
\ee
Hence, the number of sites occupied by each wave function scales with the system size as $L^u$, which also means that the localization length is infinity.

Next, we take advantage of the scale invariance of the wave function. After rescaling $r^\prime = r/b$ we discuss the problem in terms of blocks with size $b a_0$, and lose only negligible amount of short-range resonances (see Eq.~\eqref{nav}). The general tree-like structure remains the same, however hopping matrix elements $V$ become $b^3$ times larger. Since $1/V$ is the characteristic time scale, we conclude that the corresponding rescaling of time should have form $t^\prime = t/b^3$ and the problem in variables $r^\prime, t^\prime$ is the same as the original one. So, all the correlation functions should depend on combination $r^3/t$ and the dynamics is subdiffusive, $r \propto t^{1/3}$. This property was derived rigorously in Ref.~\cite{levitov1990}.

Finally, we notice, that the conception of a tree-like graph evidently fails if $N = u \ln(L/a_0) \lesssim 1$ for actual system size $L$, which means that the system size should be really large to observe delocalization effects.

\section{Discussion and conclusion}

To conclude, we show that the long-range dipolar forces can delocalize elementary excitations in three-dimensional disordered quantum magnets both in low-field and high-field gapped phases. The corresponding mechanism relies on $ \propto 1/r^3$ hopping matrix elements, which provide a tree-like network of resonances for each elementary excitation and leads to its delocalization. Renormalization group ideas are used to describe the number of states occupied by each wave function. The latter scales with the sample size as $L^u$ (exponent here is estimated with logarithmic accuracy), so the wave-functions are critical. Furthermore, their scale invariance leads to sub-diffusive $r \propto t^{1/3}$ dynamics.

Obtained results can be also important in the context of the gapless Bose glass phase. Its usual description in magnets is based on a picture of independent rare magnetically-ordered islands which are separated by non-magnetic background and do not interact with each other so their order parameters are independent. The influence of the delocalization phenomenon on this picture deserves further study.

\begin{acknowledgments}

We are grateful to A.G. Yashenkin for valuable discussions. The reported study was supported by the Foundation for the Advancement of Theoretical Physics and Mathematics ``BASIS''.

\end{acknowledgments}

\bibliography{References}

\end{document}